\documentclass[aps,pra, superscriptaddress, nofootinbib, floatfix, notitlepage]{revtex4-1}
\usepackage{bm}
\usepackage{setspace}
\usepackage{graphicx}
\usepackage{amsmath}
\usepackage{amssymb}
\usepackage{latexsym}
\usepackage{fancyhdr}
\usepackage{array}
\usepackage{color,empheq, colortbl}
\usepackage{xcolor}
\usepackage{amsfonts}
\usepackage{mathrsfs}
\usepackage{mathtools}
\usepackage{verbatim}
\usepackage{physics}
\usepackage[caption=false]{subfig}
\usepackage{natbib}
\usepackage{enumitem}
\usepackage{bbold}
\usepackage{ulem}
\usepackage[colorlinks=true,urlcolor=blue,citecolor=blue,linkcolor=blue]{hyperref}
\newcolumntype{L}[1]{>{\raggedright\let\newline\\\arraybackslash\hspace{0pt}}m{#1}}
\newcolumntype{C}[1]{>{\centering\let\newline\\\arraybackslash\hspace{0pt}}m{#1}}
\newcolumntype{R}[1]{>{\raggedleft\let\newline\\\arraybackslash\hspace{0pt}}m{#1}}
\newcolumntype{M}[1]{>{\centering\arraybackslash}m{#1}}
\newcolumntype{N}{@{}m{0pt}@{}}
\definecolor{Gray}{RGB}{220,220,220}
\definecolor{DarkGray}{RGB}{160,160,160}
\definecolor{magenta}{RGB}{255,0,255}

\begin{document}
\title{Improved entanglement indicators for optical fields
and its application in the event-ready experiment  for bright squeezed vacuum with induced non-gaussianity}

\author{Bianka Woloncewicz}
\affiliation{International Centre for Theory of Quantum Technologies, University of Gda\'nsk, 80-308 Gda\'nsk, Poland}

\author{Tamoghna Das}
\affiliation{Department of Physics, Indian Institute of Technology Kharagpur, Kharagpur-721302, India}

\author{Marek \.Zukowski}
\affiliation{International Centre for Theory of Quantum Technologies, University of Gda\'nsk, 80-308 Gda\'nsk, Poland}

\begin{abstract}
Better versions of separability conditions for four mode optical  fields, i.e. two beams with two modes of mutually orthogonal polarization are given.  Our conditions involve variances. Their meaning is intuitive and their implementation is feasible. 
Namely, if for a given quantum state  the spread of the data around its mean value is smaller than the minimal spread predicted for the set of separable states, then the  given state is entangled.
Our conditions are formulated for standard quantum  Stokes observables and  normalized Stokes observables and result to be more efficient that the previous conditions for four mode optical fields involving variances.  We test our conditions  for bright squeezed vacuum  with (and without) induced non-gaussianity obtained by addition or subtraction of photons. Also we propose a practical experimental scheme of how to generate such states for an event-ready experiment.
\end{abstract}

\maketitle
\section{Introduction}
Quantum entanglement stays beyond understanding based on classical correlations. Being an intrinsic property of Nature it underpins  fundamental phenomena and questions the idea of classical causality
Besides it finds  multiple applications in quantum optics and imaging  \cite{ZukowskiRMP, ABUimage, PETimage} quantum key distribution \cite{EkertCrypto,POPPYQKD,MAfangQKD, QKDdots}, and  quantum computing \cite{Jozsa_ent_qc_first,Gruska, QUANTcompNAT}. 
It enables quantum teleportation, superdense coding  and entanglement swapping \cite{BBCJPW, WINTERCODE, TELEPORT}.  It is then not surprising that it 
has became an intensively developing research field.

However, despite multiple breakthroughs in the understanding of entanglement its detection remains an np-hard problem \cite{GurvitzNPHard}. Most of entangled qualifiers are only sufficient entanglement conditions. Some of them turn out to be more effective than other ones for the particular quantum state. Consequently, there is still a need to construct different entanglement conditions e.g. entanglement indicators (aka witnesses) 
\cite{GUHNE_TOTH,CHRUST_SARB, HoroRMP, NAT_ENT}.

There is a multitude of entanglement conditions  tailored  for finite-dimensional scenarios i.e. qubits and qudits \cite{TerhalReview, Lewenstein00a}.  But quantum optical states of undefined number of photons also exhibit entanglement e.g.  \cite{ BRIGHTZUKU, bsv_ent,AGA_MASHA_LEUTCH}.  
With the development of experimental techniques such as  photon number resolution detectors, these theoretical concepts  can be tested in the laboratory \cite{PNR1,PNR2}, and gain  practical value in terms of use in quantum technologies.

Entanglement indicators involving correlations of intensities of optical fields were proposed in \cite{SimonBouw}. Their more efficient form  was derived in \cite{MASHA} and  
several generalizations of these were given in \cite{ZUKUSTOKES, ZUKUSTOKES1} and \cite{RYU}. In \cite{MASHA2008} it has been investigated that the variance of photon number differences in conjugate modes can also be considered as an witness of non-classicality in squeezed vacuum state.
Here, we show  a further refinement of these conditions, which leads to more efficient  entanglement indicators. 

All analyzed conditions are formulated  for standard quantum optical  observables  and  ``normalized" Stokes observables, see e.g. \cite{ZUKUSTOKES}. 
They are tailored to be  readily used for four-mode bright squeezed vacuum, which is the two beams output of  type 2 parametric down conversion (PDC) see e.g. \cite{BSV_PROP, BSV2021}.
Such states can be used in emerging quantum technologies  \cite{BSVapplic} because of its non-classical properties.  Bright squeezed vacuum (BSV), also called a ``super-singlet", exhibits perfect 
anticorrelations in polarization and perfect correlations in photon number. Still, BSV is a squeezed  state, so a gaussian state. 
We suppose that introducing non-gaussianity   to BSV might facilitate entanglement detection  with Stokes  measurement. Note that, introducing  non-gaussianity is an interesting  phenomena itself, vastly studied because of its multiple applications  see e.g. \cite{Ra2020,Takahashi2010, ENTdestilWITHnongauss, DONG} etc. Thus, it is worth to consider non-gaussian bright squeezed vacuum.
We compare (theoretically) two common techniques of introducing non-gaussianity i.e. photon addition and subtraction for bright squeezed vacuum. Our strategy is as follows. 
We  add photons to one optical beam of BSV and compare it with BSV with the same amount of photons subtracted from the another beam. W realize that  two resulting states have the same structure. Thus addition and substraction of equal amound of photons  are equivalent for bright squeezed vacuum. We propose  a feasible setup to generate BSV with added/substracted photons.
Finally, we compare the efficiency of discussed entanglement conditions for BSV and non-gaussian BSV.

\section{Improved entanglement indicators for two beams polarization entangled states} \label{Sec:Old-results}
\subsection{Standard vs. normalized Stokes operators}
Entanglement conditions for quantum optical fields require proper observables,  represented by self-adjoint operators, as prime sources of  data. 
If one is interested in polarization measurement one can use the standard Stokes operators \cite{SimonBouw} of the following form:    
\begin{equation}
    \hat \Theta_i =   \hat I_i-  \hat I_{i_\perp},
\label{OLDSTOKES}
\end{equation}
where $\hat I$ stands for intensity operator related with a given optical field  and  indices  $\{i,i_\perp\}$ denote two  orthogonal polarization degrees of freedom of that field,  related to  given three mutually unbiased  bases indexed  with $i = 1,2,3$  e.g.: $i = 1$ can stand for $\{ +45^{\circ} ,-45^{\circ} \}$ (diagonal/anti-diagonal basis), $i = 2$ is for $(\{R,L\})$ (right and  left-handed circular basis) and $i=3$  is for  $\{H,V\}$ (horizontal/vertical basis).
The  total intensity of the beam is denoted by zeroth Stokes operator: $ \hat \Theta_0 = \hat I_i + \hat I_{i_\perp}$,

From now on we will be using the model  of intensity as proportional to numbers of photons  i.e. $\hat I_i =\hat a_i^\dagger \hat a_i$, where $\hat a$ is an annihilation operator. However we emphasize that this is not the only possible model  of intensity that can be used.  Our choice is  motivated by simplicity of theoretical description. 

Normalized Stokes operators were first suggested in \cite{He_2012} in the context of Bose-Einstein condensates,  and  rediscovered for quantum optics in \cite{ZUKUSTOKES}. 
We follow the technical description given in \cite{ZUKUSTOKES}:
\begin{equation}
     \hat S_i = \hat \Pi \frac{\hat a_i^\dagger \hat a_i - \hat a_{i_\perp}^\dagger \hat a_{i_\perp}}{\hat N} \hat \Pi.
     \label{NEWSTOKES}
\end{equation}
where $\hat \Pi$, defined as $\mathbb{1} - \ket{\Omega}\bra{\Omega}$, projects out the vacuum component of a given beam, that is states of the type: 
$\hat a_i \ket{0,0} = \hat a_{i_{\perp}} \ket{0,0} = 0 $.
Note that $\hat\Pi$ is normalized zeroth Stokes operators $\hat \Pi = \hat S_0$.

In an experiment, for a single run,  the recorded values of standard and normalized Stokes observables are collected from the same set of data. 

It  was shown in \cite{ZUKUSTOKES,ZUKUSTOKES1,RYU} that normalized Stokes operators lead  to stronger entanglement conditions. 
Also, in \cite{RYU} it was shown that any linear entanglement witness  for qudits can be effortlessly  transformed into its quantum optical fields  analog  involving standard or normalized Stokes operators. 
We stress that  the discussion about standard and normalized Stokes operators  in terms of photon numbers operators takes on practical meaning as  photon number resolving detectors start to be used in the laboratory  \cite{Walmsley14, Walmsley20}.

\subsection{Separability  condition involving variances}

Let us define the  product state of the optical field.
\begin{equation}
\rho^{AB}_{\lambda} = 
f^{\dagger}_\lambda(\hat a, \hat a_{\perp})g^\dagger _\lambda(\hat b, \hat b_{\perp})\ketbra{\Omega}
f_\lambda(\hat a, \hat a_{\perp})g_\lambda(\hat b,\hat b_{\perp}),
\label{RHO-LAMBDA}
\end{equation}
where $f_\lambda(\hat a, \hat a_{\perp})$ and $g_\lambda(\hat b,\hat b_{\perp})$ are polynomial functions of annihilation operators acting on modes $a$ and $b$ related with parties $A$ and $B$ respectively. 
Mixed separable states for the studied problem have the following form
\begin{equation}
\rho^{AB}_{sep} = 
\sum_{\lambda}p_{\lambda} f^{\dagger}_\lambda(\hat a, \hat a_{\perp})g^\dagger _\lambda(\hat b, \hat b_{\perp})\ketbra{\Omega}
f_\lambda(\hat a, \hat a_{\perp})g_\lambda(\hat b,\hat b_{\perp}),
\label{RHOSEP}
\end{equation}
The  index  $\lambda$ denotes summation over elements of convex combination of product states. It might be countable or continuous.

In 2003, Simon and Bouwmeester \cite{SimonBouw} derived
an entanglement condition  based on EPR anticorrelations with standard Stokes operators

They showed that for the set of separable states, the following inequality holds 
\begin{equation}
\sum_{i=1}^3 \langle ({\hat \Theta}_{i}^A 
    +{\hat \Theta}_{i}^B )^2\rangle_{sep}  \geq 2 \langle \hat N^A + \hat N^B\rangle_{sep},    
    \label{SIMBOM}    
\end{equation}
    where $\langle . \rangle_{sep}$ is an average for any separable state $\rho^{AB}_{sep}$ \eqref{RHOSEP}.

In \cite{ZUKUSTOKES} one can find (\ref{SIMBOM}) formulated with normalized Stokes operators. It reads:
\begin{equation}
\sum_{i=1}^3 \langle ({\hat S}_{i}^A 
+{\hat S}_{i}^B )^2\rangle_{sep}  \geq  \left\langle\hat\Pi^A \frac{2}{\hat N^A}\hat\Pi^A + \hat\Pi^B\frac{2}{\hat N^B}\hat\Pi^B\right\rangle_{sep}.
\label{ZUKUSIMON}
\end{equation}
For  EPR anticorrelated states we have $\sum_{i=1}^3 \langle ({\hat S}_{i}^A 
+{\hat S}_{i}^B )^2\rangle_{EPR}=0 $. However, conditions (\ref{SIMBOM}) and (\ref{ZUKUSIMON}) are not equivalent. The latter one is more resistant to noise and losses, see \cite{ZUKUSTOKES}.

In Ref. \cite{RYU} stronger versions of entanglement indicators  (\ref{SIMBOM}) and (\ref{ZUKUSIMON}), were introduced:
\begin{equation}
\sum_{i=1}^3 \langle ({\hat \Theta}_{i}^A 
+{\hat \Theta}_{i}^B )^2\rangle_{sep}  \geq 2 \langle \hat N^A + \hat N^B\rangle_{sep} + \langle (\hat N^A -\hat N^B)^2\rangle_{sep},
\label{GENERALIMBOM}
\end{equation}
and 
\begin{equation}
\sum_{i=1}^3 \langle ({\hat S}_{i}^A 
+{\hat S}_{i}^B )^2\rangle_{sep}  \geq  \left\langle\hat\Pi^A \frac{2}{\hat N^A}\hat\Pi^A + \hat\Pi^B\frac{2}{\hat N^B}\hat\Pi^B\right\rangle_{sep} + \langle (\hat \Pi^A -\hat \Pi^B)^2\rangle_{sep}  \label{GENERALZUKSIM}.
\end{equation}

In Ref. \cite{MASHA} and \cite{ZUKOMASHA} the authors propose  to use variances of the intensities, rather than the intensities themselves. Another separability condition which is stronger than (\ref{SIMBOM}) was derived:

\begin{equation} 
\label{VAR0masha}
\begin{multlined}
\Delta \hat {\vec{ \Theta}}^{{AB}^2}_{sep} = \sum_{i=1}^3 \langle ({\hat \Theta}_{i}^A 
+{ \hat \Theta}_{i}^B )^2\rangle_{sep} - \sum_{i=1}^3\langle {\hat \Theta}_{i}^A 
+{\hat \Theta}_{i}^B \rangle^2_{sep}  \geq 2 \langle \hat N^A + \hat N^B\rangle_{sep}.
\end{multlined}
\end{equation}

In \cite{bibimaster} condition \ref{VAR0masha} for normalized Stokes operators was formulated:
\begin{equation} 
\label{VAR1masha}
\begin{multlined}
\Delta \hat{\vec{S}}^{{AB}^2}_{sep} = \sum_{i=1}^3 \langle ({\hat S}_{i}^A 
+\hat{S}_{i}^B )^2\rangle_{sep} - \sum_{i=1}^3\langle \hat{ S}_{i}^A 
+\hat{S}_{i}^B \rangle^2_{sep}  \geq 2 \left\langle \hat \Pi^A\frac{1}{\hat N^A}\hat\Pi^A + \hat \Pi^B\frac{1}{\hat N^B}\hat\Pi^B \right\rangle_{sep}.
\end{multlined}
\end{equation}

Note that, for states that exhibit perfect anticorrelations in any polarization basis ,which is the case of BSV, more precisely described in following sections,
 one has 
$\langle \hat{\Theta}_i^A + \hat{\Theta}_i^B\rangle = 0$ 
the respective above conditions (\ref{VAR0masha}) and (\ref{VAR1masha}) are reduced to
(\ref{SIMBOM}) and (\ref{ZUKUSIMON}).

\subsection{Stronger entanglement conditions based on variances}

Our aim is the further improvement of conditions (\ref{VAR0masha}) and (\ref{VAR1masha}).
In further calculations  the following  basic properties of Stokes operators will be used:  
$\sum_{i=1}^3\langle{\hat \Theta_i^X}\rangle_\lambda^2\leq \langle{\hat{N}^X}\rangle_\lambda^2$, and
$\sum_{i=1}^3\langle{\hat S_i^X}\rangle^2\leq \langle{\hat{\Pi}^X}\rangle^2$,  where $X=A,B$. 
In  this notation: $\langle{\hat{O}}\rangle_\lambda=\Tr{\hat{O}\rho^{AB}_\lambda}$, where $\hat O$ stands for an operator and $\rho^{AB}_\lambda$ is an arbitrary product state. 
For any $\rho^{AB}_{\lambda}$ it holds that  
$\langle{\hat\Theta_i^A\hat\Theta_j^B}\rangle_\lambda=
\langle{\hat\Theta_i^A}\rangle_\lambda
\langle{\hat\Theta_j^B}\rangle_\lambda$
and $\langle{\hat S_i^A \hat S_j^B}\rangle_\lambda=
\langle{\hat S_i^A}\rangle_\lambda
\langle{\hat S_j^B}\rangle_\lambda$.
Also, similarly to Pauli matrices $\{\hat\sigma_i\}_{i=x,y,z}$, which can be put in form of Pauli vector $\vec{\sigma}$  one can construct  Stokes vectors $\langle \vec{ \hat \Theta}\rangle = (\langle\hat\Theta_1\rangle, \langle\hat\Theta_2\rangle, \langle\hat\Theta_3\rangle )$ and $\langle \vec{\hat S} \rangle = (\langle\hat S_1\rangle, \langle\hat S_2\rangle, \langle\hat S_3\rangle ) $.

\subsubsection{Standard Stokes operators}
 
The variance of the Stokes vector $\Delta\vec{\hat\Theta}^{{AB}^2}$  for separable state (\ref{RHOSEP}) reads
\begin{equation}
\Delta \vec{\hat\Theta}^{{AB}^2} = \sum_{i=1}^3\Delta\hat\Theta^{A^{2}}_i +
\sum_{i=1}^3\Delta \hat\Theta^{B^{2}}_{i} + 2\sum_{i=1}^3\left(\langle\hat\Theta_i^A \hat\Theta_i^B\rangle - 
\langle \hat\Theta_i^A\rangle \langle\hat\Theta_i^B\rangle
\right).
\label{VAR1}
\end{equation}
We see that the global variance of Stokes vector for both subsystems is equal to the local variances  and global covariance of Stokes vectors for Alice and Bob that can be positive or negative.
It reads 
\begin{equation}
\begin{multlined}
\label{COV0}
\sum_{i=1}^3(\langle\hat\Theta_i^A\hat\Theta_i^B\rangle_{sep} - 
\langle\hat \Theta_i^A\rangle_{sep} \langle\hat\Theta_i^B\rangle_{sep}
){} ={}\sum_{i=1}^3\langle (\hat\Theta_i^A  - \langle \hat\Theta_i^A\rangle_{sep})(\hat\Theta_i^B - 
 \langle \hat\Theta_i^B\rangle_{sep}
\rangle_{sep}{} \\ =
{}\sum_{\lambda}\sum_{i=1}^3p_{\lambda}(\langle \hat\Theta_i^A\rangle_{\lambda}  - \langle\hat\Theta_i^A\rangle_{sep})(\langle \hat\Theta_i^B \rangle_{\lambda}- \langle\hat\Theta_i^B\rangle_{sep}
),
\end{multlined}
\end{equation}

We apply the Cauchy-Schwartz inequality to the second equality from (\ref{COV0}):
\begin{eqnarray}
&&{}\sum_{\lambda}\sum_{i=1}^3{p_{\lambda}}(\langle \hat \Theta_i^A\rangle_{\lambda}  - \langle \hat\Theta_i^A\rangle_{sep}) (\langle \hat\Theta_i^B \rangle_{\lambda}- 
\langle \hat\Theta_i^B\rangle_{sep}
){}  \nonumber  \\ 
&\leq& \sum_{\lambda}p_{\lambda}\left(\sum_{i=1}^3(\langle \hat\Theta_i^A\rangle_{\lambda} - \langle \hat\Theta_i^A\rangle_{sep})^2\right)^{\frac 12}\left(\sum_{i=1}^3(\langle \hat\Theta_i^B\rangle_{\lambda} - \langle \hat \Theta_i^B\rangle_{sep})^2\right)^{\frac 12} \label{COV1}\\ 
&=& \sum_{\lambda}\sqrt{p_{\lambda}}\left(\sum_{i=1}^3(\langle \hat\Theta_i^A\rangle_{\lambda} - \langle \hat\Theta_i^A\rangle_{sep})^2\right)^{\frac 12}\sqrt{p_{\lambda}}\left(\sum_{i=1}^3(\langle \hat\Theta_i^B\rangle_{\lambda} - \langle \hat \Theta_i^B\rangle_{sep})^2\right)^{\frac 12} \nonumber \\ 
&\leq& \left(\sum_{\lambda}p_{\lambda} \sum_{i=1}^3(\langle \hat\Theta_i^A\rangle_{\lambda} - \langle \hat\Theta_i^A\rangle_{sep})^2\right)^{\frac 12} \left(\sum_{\lambda}p_{\lambda} \sum_{i=1}^3(\langle \hat\Theta_i^B\rangle_{\lambda} - \langle \hat\Theta_i^B\rangle_{sep})^2\right)^{\frac 12} \label{COV2}.
\end{eqnarray}

Note that we have applied  Cauchy-Schwartz inequality twice: first to get the line (\ref{COV1}), with respect to the summation over $i$'s (subscripts that numerate Stokes operators) and then  the second time to get (\ref{COV2}), with respect to $\lambda$'s (i.e. the summation over a convex combination of  product states).
 	 	 
Let us observe that
\begin{equation}
\begin{multlined}
\sum_{\lambda}p_{\lambda} \sum_{i=1}^3(\langle \hat\Theta_i^X\rangle_{\lambda} - \langle \hat\Theta_i^X\rangle_{sep})^2 =\sum_{\lambda}p_{\lambda} \sum_{i=1}^3\langle \hat\Theta_i^X\rangle^2_{\lambda} - \sum_{i=1}^3\langle \hat\Theta_i^X\rangle_{sep}^2,
\end{multlined}
\label{COVEQ}
\end{equation}
where $X =A,B$.
Moreover 
\begin{equation}
\label{EST}
\sum_{\lambda}p_{\lambda} \sum_{i=1}^3\langle \hat \Theta_i^X\rangle^2_{\lambda} \leq \sum_{\lambda}p_{\lambda}\langle \hat N^X\rangle^2_{\lambda} \leq 
\sum_{\lambda}p_{\lambda}\langle \hat  N^{X^2}\rangle_{\lambda} = \langle \hat N^{X^2} \rangle_{sep}.
\end{equation}
This leads us to 
\begin{equation}
\begin{multlined}
\sum_{\lambda}p_{\lambda} \sum_{i=1}^3(\langle \hat\Theta_i^X\rangle_{\lambda} - \langle \hat\Theta_i^X\rangle_{sep})^2
\leq
\langle \hat{N}^X\rangle^2_{sep} - \sum_{i=1}^3\langle \hat\Theta_i^X\rangle_{sep}^2,
\end{multlined}
\label{COVEQ--xx}
\end{equation}

Combining (\ref{COVEQ}) and (\ref{EST}) we obtain
\begin{equation}
\begin{multlined}
{}\sum_{\lambda}\sum_{i=1}^3{p_{\lambda}}\bigg(\langle \hat\Theta_i\rangle_{\lambda}  - \langle \hat \Theta_i^A\rangle_{sep} \bigg) \bigg(\langle \hat \Theta_i^B \rangle_{\lambda}- 
\langle \hat\Theta_i^B\rangle_{sep}
\bigg) {} \\ \leq 
 \bigg(\langle \hat N^{A^2}\rangle_{sep} - \sum_{i=1}^3\langle \hat\Theta_i^A\rangle_{sep}^2\bigg)^{\frac 12}\bigg(\langle \hat N^{B^2}\rangle_{sep} - \sum_{i=1}^3\langle \hat\Theta_i^B\rangle_{sep}^2\bigg)^{\frac 12}.
\label{COVAR2}
\end{multlined}
\end{equation}

Upon involving the well-known operator equality for Stokes operators  $ \sum_{i=1}^3 \hat\Theta_i^{X^2} = \hat N^X(\hat N^X+2)$, where $X = A,B$,
the sums of local uncertainties of Stokes vector  satisfies the following
\begin{equation}
\sum_{i=1}^3\Delta \hat\Theta^{A^{2}}_{i_{sep}} +
\sum_{i=1}^3\Delta \hat\Theta^{B^{2}}_{i_{sep}} = \langle
\hat N^{A^2}\rangle_{sep}  +  \langle \hat N^{B^2}\rangle_{sep} + 2(\langle \hat N^A\rangle_{sep}  +\langle \hat N^B\rangle_{sep} ) -
\sum_{i=1}^3\langle \hat\Theta_i^A\rangle^2_{sep} - \sum_{i=1}^3\langle \hat\Theta_i^B\rangle^2_{sep}
\label{miniVAR}
\end{equation}

We combine (\ref{COVAR2}) and (\ref{miniVAR})  to estimate $\Delta \vec{\hat \Theta}^{{AB}^2}_{sep} $:
\begin{equation}\label{Bianka_expression}
\begin{multlined}
\Delta \vec{\hat \Theta}^{{AB}^2}_{sep} \geq \langle \hat N^{A^2}\rangle_{sep}  +  \langle \hat N^{B^2}\rangle_{sep} + 2(\langle  \hat N^A\rangle_{sep}  +\langle\hat N^B\rangle_{sep} ) -
\sum_{i=1}^3\langle \hat\Theta_i^A\rangle^2_{sep} - \sum_{i=1}^3\langle \hat\Theta_i^B\rangle^2_{sep}  \\ - 2\sqrt{ \bigg(\langle \hat N^{A^2}\rangle_{sep} - \sum_{i=1}^3\langle \hat\Theta_i^A\rangle^2_{sep}\bigg)\bigg(\langle N^{B^2}\rangle_{sep} - \sum_{i=1}^3\langle \hat \Theta_i^B\rangle^2_{sep}\bigg)}.
\end{multlined}
\end{equation}
The minus sign appears before the last term because we estimate the right hand side from below.
After trivial algebraic simplifications  separability condition boils down to
\begin{equation}
\begin{multlined}
\label{VARNEW}
\Delta \vec{\hat\Theta}^{{AB^2}}_{sep}  \geq 2\langle \hat N^A\rangle_{sep}  +2\langle \hat N^B\rangle_{sep}
+\bigg(\sqrt{\langle \hat N^{A^2}\rangle_{sep} -||\vec{\hat{\Theta}}^A||_{sep}^2} - 
\sqrt{\langle \hat N^{B^2}\rangle_{sep} -  ||\vec{\hat{\Theta}}^B||_{sep}^2}\bigg)^2.
\end{multlined}
\end{equation}
It is clearly seen that for states with unequal anount of photons in the beams condition \eqref{VARNEW} provide better entanglement detection than (\ref{VAR0masha}). 

\subsection{Normalized Stokes operators}
The derivation given here traces back the one given earlier, therefore its presentation will be more concise.

Consider the variance of normalized Stokes vector for the compound system of Alice and Bob:

\begin{equation}
\Delta \vec{\hat S}^{{AB}^2} = \sum_{i=1}^3\Delta\hat S^{A^{2}}_{i} +
\sum_{i=1}^3\Delta \hat S^{B^{2}}_{i} + 2\sum_{i=1}^3\left(\langle\hat S_i^A \hat S_i^B\rangle - 
\langle \hat S_i^A\rangle \langle\hat S_i^B\rangle
\right).
\label{VAR1new}
\end{equation}

The covariance term from (\ref{VAR1new}), that reads 
\begin{equation}
\begin{multlined}
\label{COV0new}
{}\sum_{i=1}^3(\langle\hat S_i^A\hat S_i^B\rangle_{sep} - 
\langle\hat S_i^A\rangle_{sep} \langle\hat S_i^B\rangle_{sep}
){} ={}\sum_{i=1}^3\langle (\hat S_i^A  - \langle \hat S_i^A\rangle_{sep})(\hat S_i^B - 
 \langle \hat S_i^B\rangle_{sep}
\rangle_{sep}{} \\ =
{}\sum_{\lambda}\sum_{i=1}^3p_{\lambda}(\langle \hat S_i^A\rangle_{\lambda}  - \langle\hat S_i^A\rangle_{sep})(\langle \hat S_i^B \rangle_{\lambda}- \langle\hat S_i^B\rangle_{sep}
){},
\end{multlined}
\end{equation}

We apply twice the Cauchy-Schwartz inequality to the last equality from (\ref{COV0new})
\begin{eqnarray}
&&{}\sum_{\lambda}\sum_{i=1}^3{p_{\lambda}}(\langle \hat S_i^A\rangle_{\lambda}  - \langle \hat S_i^A\rangle_{sep}) (\langle \hat S_i^B \rangle_{\lambda}- 
\langle \hat S_i^B\rangle_{sep}
){}  \nonumber  \\ 
&\leq& \sum_{\lambda}p_{\lambda}\left(\sum_{i=1}^3(\langle \hat S_i^A\rangle_{\lambda} - \langle \hat S_i^A\rangle_{sep})^2\right)^{\frac 12}\left(\sum_{i=1}^3(\langle \hat S_i^B\rangle_{\lambda} - \langle \hat S_i^B\rangle_{sep})^2\right)^{\frac 12} \nonumber\\ 
&\leq& \left(\sum_{\lambda}p_{\lambda} \sum_{i=1}^3(\langle \hat S_i^A\rangle_{\lambda} - \langle \hat S_i^A\rangle_{sep})^2\right)^{\frac 12} \left(\sum_{\lambda}p_{\lambda} \sum_{i=1}^3(\langle \hat S_i^B\rangle_{\lambda} - \langle S_i^B\rangle_{sep})^2\right)^{\frac 12} \nonumber.
\end{eqnarray}

Note that
\begin{equation}
\begin{multlined}
\sum_{\lambda}p_{\lambda} \sum_{i=1}^3(\langle S_i^X\rangle_{\lambda} - \langle S_i^X\rangle_{sep})^2 =
\sum_{\lambda}p_{\lambda} \sum_{i=1}^3\langle S_i^X\rangle^2_{\lambda} - \sum_{i=1}^3\langle S_i^X\rangle_{sep}^2.
\end{multlined}
\label{COVEQnew}
\end{equation}
For normalized Stokes operators  we have
$\sum_{\lambda}p_{\lambda} \sum_{i=1}^3\langle S_i^X\rangle^2_{\lambda} \leq \sum_{\lambda}p_{\lambda}\langle \hat\Pi^X\rangle_{\lambda} = \langle \hat{\Pi}^X\rangle$. 
where we used the fact that $\hat{\Pi}^X$ is a projector (and again $X = A,B$). Thus,
\begin{equation}
\begin{multlined}
{}\sum_{\lambda}\sum_{i=1}^3{p_{\lambda}}\bigg(\langle S_i^A\rangle_{\lambda}  - \langle S_i^A\rangle_{sep} \bigg) \bigg(\langle S_i^B \rangle_{\lambda}- 
\langle S_i^B\rangle_{sep}
\bigg) {} \\ \leq 
 \bigg(\langle \hat \Pi^{A}\rangle_{sep} - ||\vec{\hat S}^A||^2_{sep}\bigg)^{\frac 12}\bigg(\langle \hat \Pi^{B}\rangle_{sep} - ||\vec{\hat S}^B||_{sep}^2\bigg)^{\frac 12}.
\label{COV2new}
\end{multlined}
\end{equation}

Let us analyze the local  uncertainties  of (\ref{VAR1new}):
\begin{equation}
\begin{multlined}
\Delta \hat{\vec{S}}^{A^{2}}_{{sep}} +
\Delta \hat{\vec S}^{B^{2}}_{{sep}}  \geq  2( \langle\hat{\Pi}^A\frac{1}{\hat N^A}\hat\Pi^A\rangle_{sep} + 
\langle\hat{\Pi}^B\frac{1}{\hat N^B}\hat\Pi^B\rangle_{sep})
\label{miniVARnew},
\end{multlined}
\end{equation}
where we used the operator equality given in \cite{ZUKUSTOKES}:
$\sum_{i=1}^3\hat S_i^{X} = \hat{\Pi}^X  +\hat \Pi^X\frac{2}{\hat{N}^X}\hat \Pi^X$. Combining (\ref{COV2new}) with (\ref{miniVARnew}) after simplifications we obtain
\begin{equation}
\label{NEWVAR_NEWSTOKES}
\begin{multlined}
\Delta \vec{\hat{S}}^{{AB}^2}_{sep}   \geq 2(\langle\hat\Pi^A \frac{1}{\hat N^A}\hat\Pi^A\rangle_{sep}  +\langle\hat\Pi^B \frac{1}{\hat N^B}\hat\Pi^B \rangle_{sep} )
+\bigg(\sqrt{\langle\hat\Pi^{A}\rangle_{sep} - || \vec{\hat{S}}^A||^2_{sep}} - 
\sqrt{\langle\hat\Pi^{B}\rangle_{sep} - ||\vec{\hat{S}}^B||^2_{sep}}\bigg)^2.
\end{multlined}
\end{equation}

Thus we get a tighter constraint on separability.

\section{Bright squeezed vacuum with added and subtracted photons}

Bright squeezed vacuum (BSV) consists of  two optical beams (directions) in which photon pairs are emitted. Each beam contains two  optical modes carrying  mutually perpendicular polarizations (we choose:  horizontal-$H$ and vertical-$V$). It reads:
\begin{equation}
\ket{BSV}=
\frac{1}{\cosh^2 \Gamma}\sum_{n=0}^\infty\tanh^n \Gamma\sum_{r=0}^n(-1)^m\ket{{(n-r)}_{a_H},{r}_{a_V},{r}_{b_{H}},{(n-r)}_{b_{V}}}.
\label{BSV_HERO}
\end{equation}
where $\Gamma$ is the amplification gain.
Subscripts $a$ and $b$ stand for  two beams that reach two observers $A$ and $B$.

From the formula \eqref{BSV_HERO} we see that   BSV exhibit perfect correlations in the numbers of photons and perfect anti-corralations of polarization modes $\{H,V\}$. 
Moreover  BSV is rotationally invariant with respect to the same rotations of both observers, and thus its form remains unchanged in any other polarization basis $\{i, i_{\perp}\}$.
Thus, because of  its perfect EPR correlations independent of chosen  polarization basis and the same number of photons in the two beams  all   entanglement  conditions concidered in this paper  are equivalent. 
In order to observe  the possible advantage of  our conditions (\ref{VARNEW}) and (\ref{NEWVAR_NEWSTOKES})  we need  to introduce an asymmetry in the number of photons in the beams. 

Adding and subtracting photons are two well-known techniques to induce non-gaussianity into photon statistics \cite{KIM_SUB_ADD}, \cite{ZAV_SINGLE_PHOTON}.
Both  processes are  feasible in the laboratory \cite{SUB_ADD_STAT}, \cite{KITTEN} \cite{SUBSTRACTION_TREPS}, but subtraction of photons is experimentally easier: it can be realised using a  beamsplitter  with high transmitivity and a photon number resolving detector behind one the beamspliter's outputs \cite{GRANGIER}. Photon addition is more challenging.  It can be  achieved by feeding the photon in the respective mode to the input of a parametric amplifier and detecting  single  photons  in the idler output of parametric process \cite{Zavatta660}, \cite{ZAV_SUB_ADD}.

\subsection{Equivalence of photon added and photon subtracted BSV}
\label{sec:addeqsub}
 
Let us first analyse the processes of adding and subtracting photons to BSV from  purely theoretical point of view.
We shall show their equivalence. 

Bright squeezed vacuum with $m_1$ photons added in mode $a_H$ and $m_2$ photons added  in mode $a_V$ reads
\begin{eqnarray}
\nonumber
\ket{BSV^{add}}&=& \frac{1}{\sqrt{N^{add}}} (\hat a_H^\dagger)^{m_1} (\hat a_V^\dagger)^{m_2} \ket{BSV}\\
&=& \frac{1}{\sqrt{N^{add}}}  \sum_{n = 0}^\infty \tanh^n{\Gamma} \sum_{r =0}^n (-1)^r \sqrt{\frac{(n-r+m_1)!}{(n-r)!}}\sqrt{\frac{(r+m_2)!}{r!}} \nonumber \\
&& \hspace{1.7in} \ket{(n-r+m_1)_{a_H}, (r+m_2)_{a_V}, r_{b_H},(n-r)_{b_V}}, \label{eq:added} 
\end{eqnarray}
where the normalization factor $N^{add}$ is given by 
\begin{eqnarray}
 N^{add} = \sum_{n = 0}^\infty \tanh^{2n}{\Gamma} \sum_{r =0}^n  \frac{(n-r+m_1)!}{(n-r)!}\frac{(r+m_2)!}{(r)!} =   m_1! m_2! (\cosh^2{\Gamma})^{m_1 + m_2 + 2}.
\end{eqnarray}

Now consider  the subtraction of 
the same amount of $m_H$ and $m_V$ photons  from the second optical beam $b$. 
This time we subtract   
 $m_1$ photons  from mode $b_{V}$
 and $m_2$ photons from the mode $b_H$  i.e.   the same amount of photons  that in the previous example were  added  to  modes $a_{H}$ and $a_V$.
We get:
\begin{eqnarray}\ket{BSV^{sub}} &=& \frac{1}{\sqrt{N^{sub}}} (\hat b_H)^{m_2} (\hat b_V)^{m_1} \ket{BSV} \nonumber \\
&=& \frac{1}{\sqrt{N^{sub}}}  \sum_{n = m_1+m_2}^\infty \tanh^n{\Gamma} \sum_{r =m_2}^{n-m_1} (-1)^r \sqrt{\frac{(n-r)!}{(n-r - m_1)!}}\sqrt{\frac{(r)!}{(r-m_2)!}} \nonumber \\
& & \hspace{1.6in} \ket{(n-r)_{a_H}, r_{a_V}, (r - m_2)_{b_H},(n-r - m_1)_{b_V}} \label{eq:sub_BSV}
\end{eqnarray}
and $N^{sub}$ is given by: 
\begin{eqnarray}
 N^{sub} = \sum_{n = m_1+m_2}^\infty \tanh^{2n}{\Gamma} \sum_{r =m_2}^{n-m_1} \frac{(n-r)!}{(n-r - m_1)!}\frac{r!}{(r-m_2)!}. 
\end{eqnarray}

We perform the following change of variables: $n-m_1-m_2 = n'$ and $r - m_2 = r'$. As the result we get: $n = n'+ m_1+ m_2$ and $r = r'+ m_2$. Thus, $\sum_{n = m_1 +m_2}^\infty \rightarrow \sum_{n' = 0}^\infty$ and $\sum_{r =m_2}^{n-m_1} \rightarrow \sum_{r' =0}^{n'}$. Applying these changes to formula \eqref{eq:sub_BSV} we obtain
\begin{eqnarray}
&&\ket{BSV^{sub}}\label{eq:added_alias} \nonumber \\
&=& \frac{1}{\sqrt{N^{sub}}}  \sum_{n' = 0}^\infty (\tanh{\Gamma})^{n'+m_1 + m_2} \sum_{r' =0}^n (-1)^{r'+ m_2} \sqrt{\frac{(n'-r' + m_1)!}{(n'-r')!}}\sqrt{\frac{(r+m_2)!}{r!}}  \nonumber \\
&& \hspace{2.5in} \ket{(n'-r'+m_1)_{a_H}, (r'+m_2)_{a_V}, r'_{b_H},(n'-r')_{b_V}}
\nonumber \\
&=& \frac{(-1)^{m_2}}{\sqrt{N^{add}}}  \sum_{n = 0}^\infty \tanh^n{\Gamma} \sum_{r =0}^n (-1)^r \sqrt{\frac{(n'-r' + m_1)!}{(n'-r')!}}\sqrt{\frac{(r+m_2)!}{r!}}  \nonumber \\
&& \hspace{2.5in} \ket{(n'-r'+m_1)_{a_H}, (r'+m_2)_{a_V}, r'_{b_H},(n'-r')_{b_V}} \nonumber\\ 
&=&\ket{BSV^{add}},
\end{eqnarray}
where in the second equality, we introduced that $N^{sub} = \tanh^{2(m_1 + m_2)}\Gamma N^{add}$.  

Hence, we showed that $\ket{BSV^{sub}}$   and $\ket{BSV^{add}}$
are equivalent
as they differ only  by the 
global phase factor $(-1)^{m_2}$.
This is one of the consequences  of the   symmetry of BSV state and it is very intuitive. 
Note that
all the ``kets" of $\ket{BSV^{add}}$
can be as well considered as ``kets" belonging to BSV with  photons substracted in the second  beam.  For example, the ``ket" $\ket{2_{a_H}, 0_{a_V}, 0_{b_H}, 1_{b_V}}$ can be seen  either as $\ket{1_{a_H}, 0_{a_V}, 0_{b_H}, 1_{b_V}}$  with one photon $m_1=1$ added in mode $a_H$ or as $\ket{2_{a_H}, 0_{a_V}, 0_{b_H}, 2_{b_V}}$ with one ($m_1=1$) photon subtracted  from the  mode $b_V$.

\subsection{Experimental setup for generating BSV with induced non-gaussianity}

In  order to get BSV with added or substracted photons, it is enough to subtract photons with polarizing beamsplitters. 
First we show that  applying  
annihilation operator $m$ times on the given state  can be  physically realizable by  subtracting $m$ photons with highly transmissive beamsplitter. 
Let us analyze these processes from more physical perspective starting by the analysis of photon subtraction from a single mode.

\subsubsection{Photon subtraction from a single mode state}
 
Consider a a beamsplitter $BS(T)$ with arbitrary transmitivity $T > 0$. The input arms are denoted with  $a$  and  $b$. The output ports are $c$ and $d$ (see figure \ref{fig:schematic_sub}(a)).  The relation between the creation operators assigned to the  input and the output modes is given by the unitary transformation
\begin{eqnarray}
 \hat a^\dagger = \sqrt{T} ~\hat c^\dagger + i \sqrt{1 - T} ~\hat d^\dagger, \\
 \hat b^\dagger = i \sqrt{1 - T} ~\hat c^\dagger + \sqrt{ T} ~\hat d^\dagger.
\end{eqnarray}

Assume $\ket{\psi}_a = \ket{n}_a$  with $n > 0$ is fed in the mode $a$ and there is  no optical input in the mode $b$. After passing through the beamsplitter  the state yields 
\begin{eqnarray}
\label{BS_def}
\nonumber
 \ket{n_a,\Omega_b} &=& \frac{1}{\sqrt{n!}} (\hat a^\dagger)^n \ket{\Omega_{a,b}} = \frac{1}{\sqrt{n!}} (\sqrt{T} ~\hat c^\dagger + i \sqrt{1 - T} ~\hat d^\dagger)^n \ket{\Omega_{c,d}} \\
\nonumber
 &=& \frac{1}{\sqrt{n!}} \sum_{s = 0}^n \binom{n}{s} (\sqrt{T})^{n-s} (i \sqrt{1 - T})^s (\hat c^\dagger)^{n-s} (\hat d^\dagger)^s \ket{\Omega_{c,d}} \\
\nonumber
 &=& \frac{1}{\sqrt{n!}} \sum_{s = 0}^n \binom{n}{s} (\sqrt{T})^{n-s} (i \sqrt{1 - T})^s \sqrt{(n-s)!s!}  \ket{(n-s)_c, {s}_d}, \\
 &=& \sum_{s = 0}^n \sqrt{\frac{n!}{(n-s)!s!}} (\sqrt{T})^{n-s} (i \sqrt{1 - T})^s   \ket{(n-s)_c,{s}_d} = \ket{\varphi}_{cd}.
\end{eqnarray}

Note that if a photon number resolving detector  placed in the mode  $d$  detects $m$-photon state $\ket{m}_d$, we know that  $m$ photons were subtracted from the mode $c$ i.e. we  know that we have  $\ket{n-m}_c$ in the mode $c$  (see figure \ref{fig:schematic_sub}(a)).  We can then use $\ket{n-m}_c$ for further measurements  that  we start  to run when the detector detects $m$ photons. Thus, such experiments are event-ready.   

Now consider a state $\ket{\psi}_a$ that is a linear combination of  photon number states  i.e. $\ket{\psi} = \sum_{n = 0}^\infty q_n \ket{n}$, with $\sum_{n = 0}^\infty |q_n|^2 = 1$. Again, we send  it on the beamsplitter (Fig. \ref{fig:schematic_sub}(a))  and we place a detector what is going to measure  $m$-photons  in  the  output  mode $d$.  The  state  reads
\begin{eqnarray}
\nonumber
\ket{\psi}_a \ket{\Omega}_b =
\sum_{n=0}^\infty q_n \ket{n_a,\Omega_b} &=& \sum_n^\infty  \frac{q_n}{\sqrt{n!}} (\hat a^\dagger)^n \ket{\Omega_{a,b}} = \sum_{n=0}^\infty  \frac{q_n}{\sqrt{n!}} (\sqrt{T} ~\hat c^\dagger + i \sqrt{1 - T} ~\hat d^\dagger)^n \ket{\Omega_{c,d}} \\
\nonumber
&=& \sum_{n=0}^\infty  \frac{q_n}{\sqrt{n!}} \sum_{s = 0}^n \binom{n}{s} (\sqrt{T})^{n-s} (i \sqrt{1 - T})^s (\hat c^\dagger)^{n-s} (\hat d^\dagger)^s \ket{\Omega_{c,d}} \\ 
&=& \sum_{n=0}^\infty  q_n \sum_{s = 0}^n \sqrt{\frac{n!}{(n-s)!s!}} (\sqrt{T})^{n-s} (i \sqrt{1 - T})^s   \ket{(n-s)_c,s_d}.
\end{eqnarray}

Thus after $m$ photons are detected  the  state with $m$ photons subtracted is given by  
\begin{eqnarray}\label{eq:psi_sub}
\ket{\psi^{sub}} = \frac{1}{\sqrt{N^{sub}}} \sum_{n = m}^{\infty} q_n \sqrt{\frac{n!}{(n-m)!}} (\sqrt{T})^{n-m} \ket{(n-m)_c}, 
\end{eqnarray}
where $N^{sub}$ is the normalization constant: $N^{sub} = \sum_{n = m}^\infty |q_n|^2 \frac{n!}{(n-m)!} T^{n-m}$. 

If the beamsplitter is highly transmissive     we get 
\begin{eqnarray}
\lim_{T \rightarrow 1} \ket{\psi^{sub}}  =
\frac{1}{\sqrt{\tilde{N}^{sub}}} \sum_{n = m}^\infty q_n \sqrt{\frac{n!}{(n-m)!}} \ket{(n-m)} = \frac{a^m \ket{\psi}}{\sqrt{\tilde{N}^{sub}}},
\label{TRANS_APPROX}
\end{eqnarray}
where $\tilde{N}^{sub} = \sum_{n = m}^\infty |q_n|^2 \frac{n!}{(n-m)!}$, and the transmitted mode can be approximated to $c \approx a$.

Thus, subtracting $m$ photons with highly transmissive beamsplitter can be approximated by applying  
annihilation operator $m$ times. 
The  probability of subtraction decreases as more photons are  subtracted, and for $m>>1$ it becomes vanishingly small. 
 
\subsubsection{Bright squeezed vacuum with added or subtracted photons.}

\begin{figure}
    \centering
    \includegraphics[width= 0.8
\columnwidth]{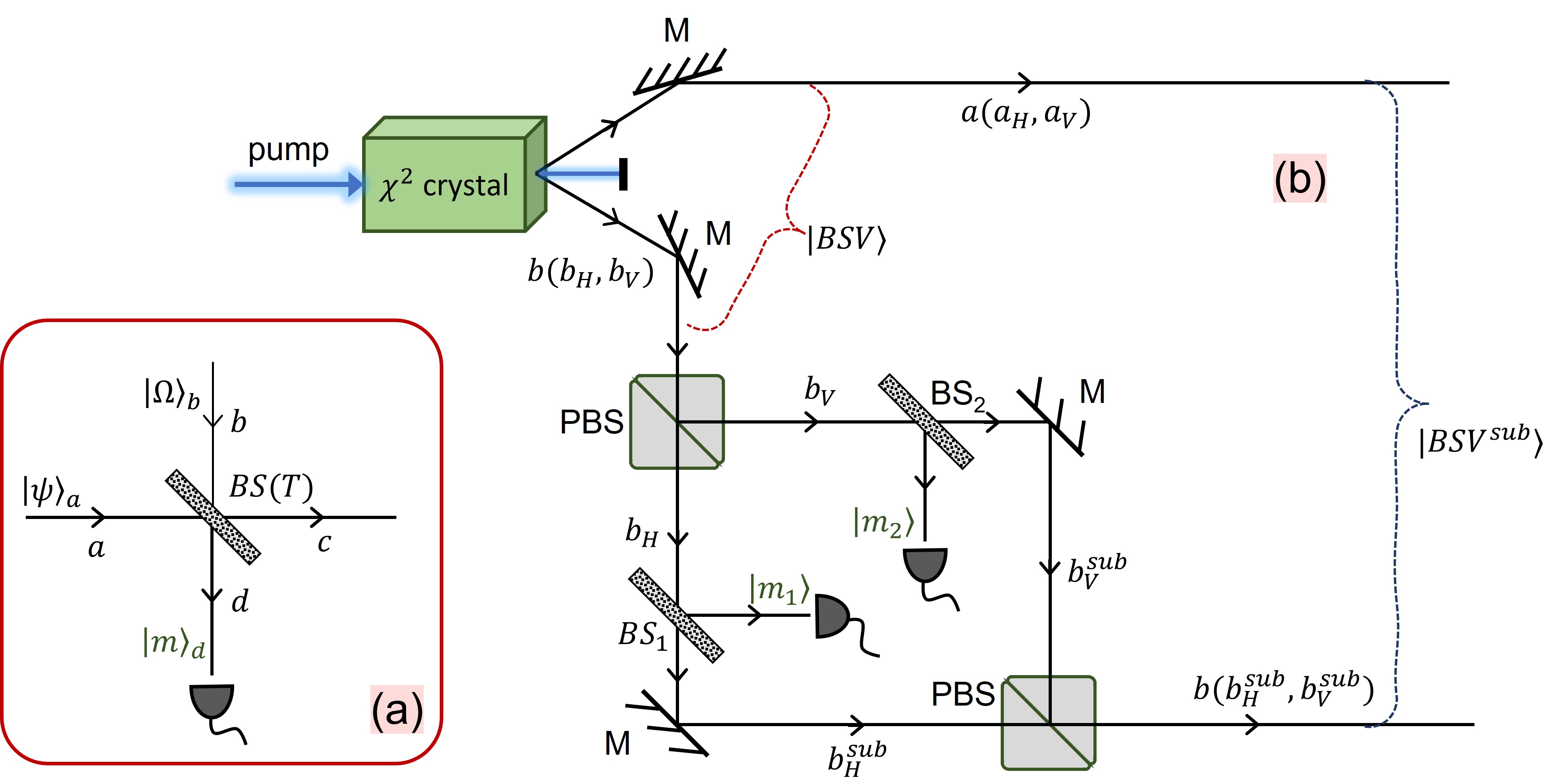}
    \caption{Schematic of photon substraction. Panel (a) shows a
    subtraction of $m$ photons from an arbitrary state $\ket{\psi}$ feeding  the input arm $a$ of a beamsplitter of transmitivity $T$.  On the reflected output mode $d$ we place a photon number resolving  detector. If  $m$ photons are registered then the transmitted mode $c$ contains a state with $m$ photons subtracted. Panel (b): schematic setup
to generate non-gaussianity in bright squeezed vacuum with the amplification gain $\Gamma'$.  Our aim  is to subtract 
$m_1$ photons from mode $b_H$ and $m_2$ photons from mode $b_V$ of BSV  The beam $b$ impinges on the polarizing beamsplitter (PBS) so that horizontally and vertically polarized photons from $b$ becomes spatially separated,
    The  beams $b_H$ and $b_V$ pass through the  beamsplitters $BS_1$ and $BS_2$. 
     Once  $m_1$  and $m_2$  photons  are detected in the reflected  output beams, photon subtracted states in modes
     $b_H^{sub}$ and  $b_V^{sub}$ generated. Note that such a preparation of BSV leads to to possibility of having an event-ready experiment. 
 }
    \label{fig:schematic_sub}
\end{figure}

Fig. \ref{fig:schematic_sub}(b) shows how to induce non-gaussianity in BSV. Bright squeezed vacuum is generated via PDC procces. amplification gain is $\Gamma'$. The
The beam $b$  passes through a polarizing beamsplitter (PBS) that separates spatially modes $b_H$ and $b_V$. 
The  mode $b_H$ impinges on a beamsplitter $BS_1$ while the mode $b_V$  passes through $BS_2$. Both beamsplitters $BS_1$ and $BS_2$ have the same transmissivity $T$. We place photon numbers resolving detectors on the optical paths of reflected beams. 
If  $m_1$ and $m_2$ photons are detected we know that the  transmitted beams $b_H^{sub}$ and  $b_V^{sub}$ have  $m_1$ and $m_2$ photons subtracted.  Finally beams $b_H^{sub}$  and $b_V^{sub}$ are recombined into $b(b_H^{sub},b_V^{sub})$ with subtracted photons. The beam $a$ remains unchanged. The output state reads
\begin{eqnarray}
&&\ket{BSV_T^{sub}(\Gamma')} \label{eq:sub_BSV_BS} \\
&=& \frac{1}{\sqrt{N_T^{sub}(\Gamma')}}  \sum_{n = m_1+m_2}^\infty \tanh^n{\Gamma'} \sum_{r =m_1}^{n-m_2} (-1)^r \sqrt{\frac{(n-r)!}{(n-r - m_2)!}}\sqrt{\frac{r!}{(r-m_1)!}} (\sqrt{T})^{n - m_1 - m_2} \nonumber \\
&&  \ket{(n-r)_{a_H}, r_{a_V}, (r - m_1)_{b_H},(n-r - m_2)_{b_V}},~~~~~
\end{eqnarray}
where $N_T^{sub}(\Gamma')$ is the normalization constant depending  on $T$ given by 
\begin{eqnarray}
 N_T^{sub}(\Gamma') = \sum_{n = m_1+m_2}^\infty \tanh^{2n}{\Gamma'} \sum_{r =m_1}^{n-m_2} \frac{(n-r)!}{(n-r - m_2)!}\frac{r!}{(r-m_V)!} T^{n - m_1 - m_2}. 
\end{eqnarray}

Let us replace $m_1 \to m_V$, $m_2 \to m_H$, and $\tanh{\Gamma'} \to \frac{1}{\sqrt{T}}\tanh{\Gamma}$. Then $N_T^{sub}(\Gamma') T^{  m_1 + m_2}= N^{sub}(\Gamma) $ and in consequence, \eqref{eq:sub_BSV_BS} with subtracted photons  and amplification gain $\Gamma'$ becomes (\ref{eq:added}) with the aplification gain $\Gamma$ and added photons. 
Hence the conclusion that  in order to obtain the photon added BSV($\Gamma$), we need to create photon subtracted  BSV ($\Gamma'$)  such that  $\tanh{\Gamma} = \tanh{\Gamma'} \sqrt{T}$.   
The use of  BSV with unequal number of photons in the beams may facilities the detection of entanglement. Moreover, it has the adventage of an event-ready experiment. The state with $m$ photons added or subtracted  is signalized by the detection of these $m$ photons (a detector click).

\section{Comparison  of  entanglement indicators involving variances}
\label{COMPARISON}

We compare our  condition \eqref{VARNEW}  with the previous variance condition \eqref{VAR0masha} from \cite{MASHA2008}  for BSV with (and without) substracted photons.   
For the simplicity of further comparison  
we give all separability conditions in the form of
$LHS_{sep} \geq RHS_{\hat X_{sep}}$ where $LHS$ - the left hand side - remains the same for all the conditions (in its standard or normalized version). The right hand side $RHS_{\hat X}$ changes according to the given condition indexed with $\hat X$  e.g. $\hat X = {\Delta(\hat \Theta^{AB})^2}$ denotes the variance condition \eqref{VAR0masha}  for standard Stokes operators  etc.  

The left hand side for standard Stokes operators is:
$LHS= \langle (\hat {\vec{ \Theta}}^A + \hat {\vec{ \Theta}}^B)^2\rangle$. 
The  different right hand sides are given by: 
\begin{itemize}
\item{$\label{RHS_VAR}
    RHS_{\Delta(\hat \Theta^{AB})^2} =
2\langle \hat N^A + \hat N^B\rangle
+ \sum_{i=1}^3 \langle ({\hat \Theta}_{i}^A 
+{\hat \Theta}_{i}^B )\rangle^2$ for
 \eqref{VAR0masha}} 
\item{$RHS_{\Delta(\hat \Theta^{AB})^2_{{new}}}
  = RHS_{\Delta(\hat \Theta^{AB})^2} + \bigg(\sqrt{\langle \hat N^{A^2}\rangle -\langle \hat{\vec{\Theta}}^A\rangle^2} - 
\sqrt{\langle \hat N^{B^2}\rangle - \langle \hat{\vec{\Theta}}^B\rangle^2}\bigg)^2  
$
for our (\ref{VARNEW})}
\end{itemize}

We use the same reasoning for normalized Stokes operators.
  
 We define $\kappa= LHS - RHS_{\hat X}$. We  For any separable state $\kappa_{sep} \geq 0$. 
If  $\kappa < 0$, the  state in question is  entangled.
The more negative $\kappa$ is, the  more efficient is  detection of entanglement.

\begin{figure}
	\centering
\includegraphics[width= 0.95
\columnwidth]{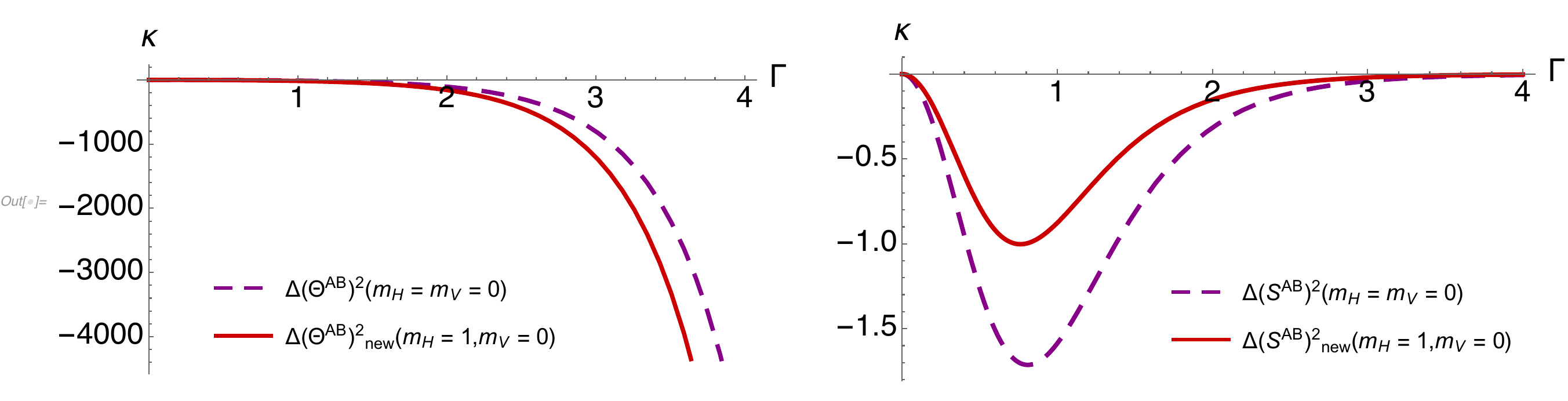}
	\caption{
 \label{comcom}
Plots of c$\kappa$ in function of amplification gain $(\Gamma)$. 
On the left: 
comparison of conditions  (\ref{VAR0masha}) applied for BSV and (\ref{VARNEW}) for non-gaussian BSV  (with one photon horizontally polarized $m_H=1$ subtracted) for standard Stokes operators. On the right analogous comparison of  (\ref{VAR1masha}) and  (\ref{NEWVAR_NEWSTOKES}). 
The photon addition combined with the use of our condition  helps in entanglement detection only for standard Stokes operators. }
\end{figure}

Let us first analyze if using our modified  entanglement condition applied to  BSV with one horizontally polarized  photon subtracted   has any adventage at all over using  its  "basic" version i.e.  the  condition involving variances  from \cite{MASHA2008} and BSV. Fig. \ref{comcom} shows $\kappa(\Gamma)$ for these scenarios for standard (see the subfigure on the left) and normalized Stokes operators (the subfigure on the right).
On the left  side of fig. \ref{comcom} we find   $\kappa(\Gamma)$ for condition (\ref{VARNEW}) applied to non-gaussian BSV put together with $\kappa(\Gamma)$ for condition  \ref{VAR0masha} and ``regular"  BSV for standard Stokes operators. Our modified scenario leads to stronger detection of entanglement.
Then, on the right  side of \ref{comcom} we have similar plots for normalized Stokes operators (we compare conditions \ref{VAR1masha} for BSV and \ref{NEWVAR_NEWSTOKES} for non-gaussain BSV).
Concerning normalized Stokes operators  for low values of $\Gamma$ it is more advantageous to use the ``basic" scenario with  \eqref{VAR1masha} than to add one photon from BSV and use our  condition (\ref{NEWVAR_NEWSTOKES}).
 
In the main text we  concentrate on standard Stokes operators for which the  analytical expressions for $LHS$ and all $RHS_{\hat X}$ are given in appendix \ref{FORMULAS}.
The corresponding    plots and formulas for normalized Stokes operators
can be found in appendix \ref{NEW_TAB}.

\begin{figure}
	\centering
\includegraphics[width= 0.95
\columnwidth]{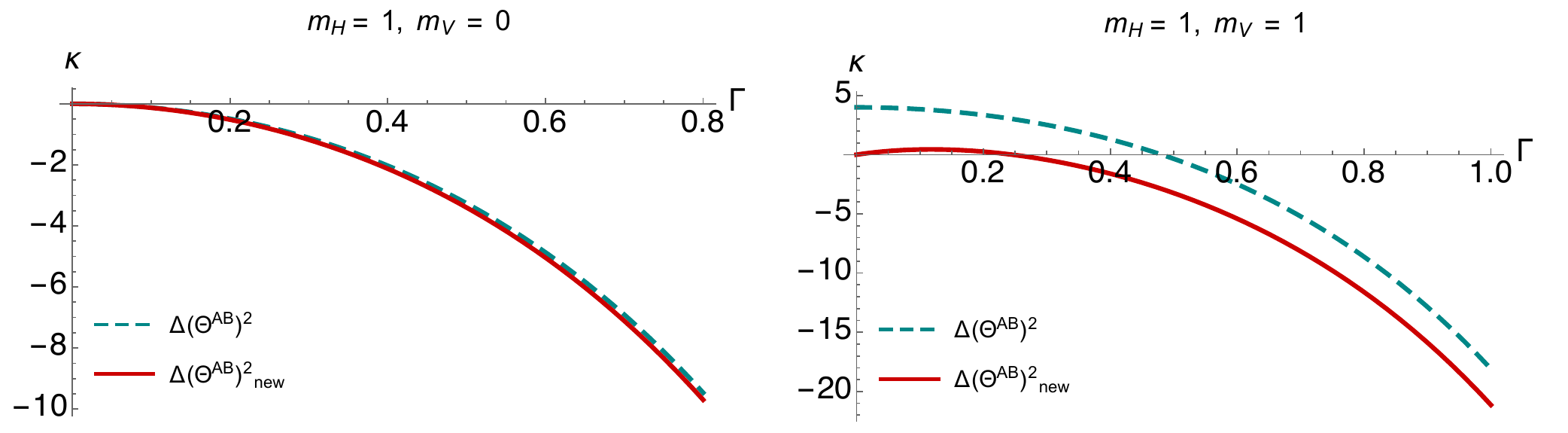}
	\caption{
 \label{2modeaddStokes}
Plots of $\kappa$ in function of amplification gain $(\Gamma)$.
for  conditions  (\ref{VAR0masha}) and (\ref{VARNEW})  for BSV with $m_H$ and $m_V \in  \{0,1\}$ photons added. Note the adventage of our condition (\ref{VARNEW}) - entanglement can be observed for smaller value of amplification gain $(\Gamma \approx 0.2)$ than for  (\ref{VAR0masha}) }
\end{figure}

Now let us check the behaviour of analyzed entanglement conditions when they are both applied to  BSV  with different numbers added photons. 
Fig.  \ref{2modeaddStokes}  shows the  values of $\kappa$ in  function of the amplification gain $\Gamma$ for condition (\ref{VAR0masha}) and (\ref{VARNEW}). 
We checked that as we increase the difference of total number of photons  between the  beams (by subtracting  $m_H$ and or $m_V$ to  one beam), the additional non-negative term in $RHS_{\Delta(\hat \Theta^{AB})^2_{{new}}}$ of 
(\ref{VARNEW}) gets more impact on lowering the value of $\kappa$.  
As expected we noted that condition  \eqref{VAR0masha} is unable to capture entanglement if we  subtract  (or add) photons from BSV for low values of $\Gamma$, We did not put results for  subtracting more than  2 photons in the beam because the obtained results  are conceptually  similar to the cases  $m_H$ and $m_V \in  \{0,1\}$.

\section{Conclusions}

More optimal conditions for optical fields tailored for states with undefined number of  photons were proposed. Our conditions, based on polarization measurement, involve variances and hence they have clear physical interpretation - the spread of data around the mean value define the nature of correlations. If for the given state the spread of the data results smaller than it is assumed  for separable states, the given state is entangled. 
We compared the efficiency of entanglement detection with our condition versus other known entanglement condition based on variances for bright squeezed vacuum and bright squeezed vacuum with added photons. We also analyzed the equivalency of adding and subtracting photons in bright squeezed vacuum and proposed  an experimental scheme to perform such experiments. Our scheme  leads  event-ready detection of entanglement for BSV.
We have shown  that by  applying our conditions to  non-gaussian bright squeezed vacuum,   we can obtain stronger entanglement detection. However, that statement refers only to standard Stokes operators. For normalized Stokes operators adding photons to BSV and using our condition does not improve entanglement detection compared to other studies cases.   

\appendix
\section{Explicit formulas used in the  comparison of entanglement conditions for  standard Stokes operators}
\label{FORMULAS}

We give the analytical form of the formulas  for $LHS$ and $RHS_{\hat X}$  for  (\ref{VAR0masha}) and (\ref{VARNEW}). 
The left hand side, the same for all conditions, boils down to 
\begin{equation}
LHS = (m_H + m_V + 1)^2 - 1.
\end{equation}
Note that $LHS$ does not depend on $\Gamma$, which is not surprising because for if no photon are added the $LHS = 0$, due to perfect EPR anticorrelations of BSV.
Then, to calculate the  right hand sides $(RHS_{\hat X})$ we need the following components:
\begin{equation}
\label{Nb}
 \langle \hat N^B \rangle_{add} = (m_H + m_V + 2)\sinh ^2 \Gamma,
\end{equation}
\begin{equation}
\label{Na^2} 
\langle \hat N^{A^2}
\rangle_{add} = 
\sinh ^2\Gamma  (m_H+m_V+2) \left(1 + \sinh ^2 \Gamma (m_H + m_V+3)\right),
\end{equation}
\begin{equation}
\label{thetaAB}
 \langle \hat \Theta_3^A \hat \Theta_3^B \rangle_{add} = -\frac{1}{2} (1 + m_H) (1 + m_V) \sinh ^2(2 \Gamma ),
\end{equation}
\begin{equation}
\label{NANBave}
 \langle \hat N^A \hat N^B \rangle_{add} =  (m_H + m_V + 2) \cosh ^2\Gamma \sinh ^2\Gamma (1 + m_H + m_V + 2 \tanh^2 \Gamma).
 \end{equation} 
Note that after introducing extra photons BSV ceases to be rotationally invariant (with respect to the same rotations of both observers). We still have: $\langle \hat \Theta_1^X \rangle_{add} = \langle \hat \Theta_2^X \rangle_{add} = 0$, for $X = A, B$, but 
\begin{equation}
\label{theta}
 \langle \hat \Theta_3^A \rangle_{add}  - (m_H - m_V) \sinh^2{\Gamma}.
\end{equation} 

The formulas listed above were used to obtain plots on  fig. \ref{comcom} and fig.\ref{2modeaddStokes} in the main text.

\section{Comparison of entanglement conditions for normalized Stokes operators and conditions involving variances.}
\label{NEW_TAB}
\begin{figure}[h]
	\centering
\includegraphics[width= 0.8
\columnwidth]{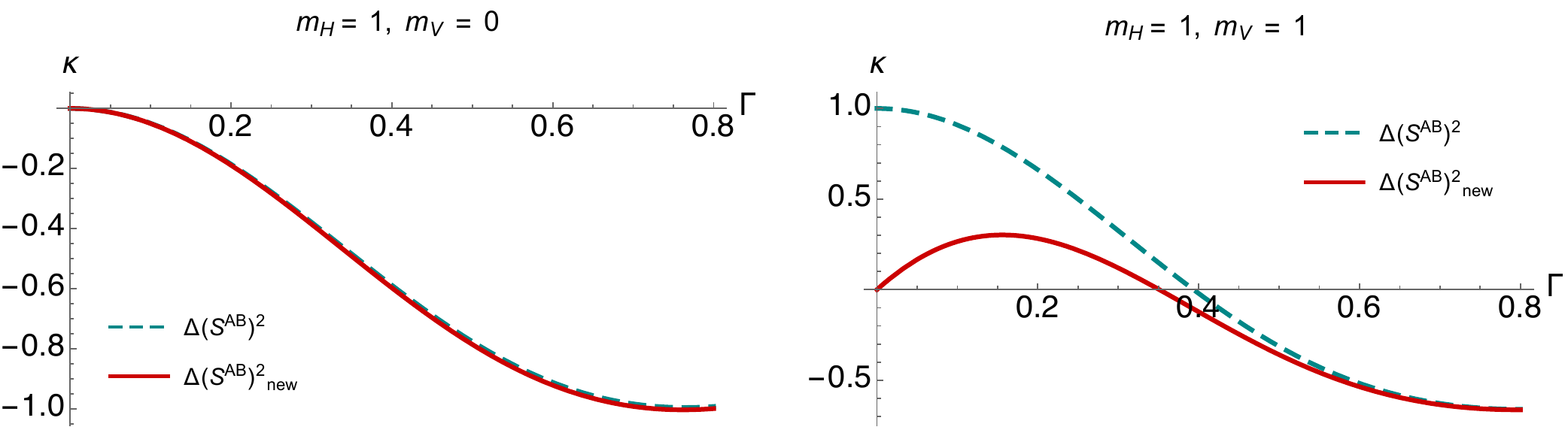}
\caption{\label{2modeaddNStokes}
The plots of $\kappa(\Gamma)$ for  
(\ref{VAR1masha}) and \ref{NEWVAR_NEWSTOKES} for different amount of added photons $m_{H_{(V)}}$.}
\end{figure}

We repeat the reasoning form the begining of section \ref{COMPARISON}  in the main text  for normalized Stokes parameters. We have : 
\begin{equation}
LHS_{\hat S^{AB}}= \sum_{i=1}^3 \langle ({\hat S}_{i}^A 
+{\hat S}_{i}^B )^2\rangle, 
\end{equation}
as well as:
\begin{itemize}
\item     $RHS_{\Delta(\hat S^{AB})^2} =
2\langle \hat \Pi^A\frac{1}{\hat N^A}\hat \Pi^A + 
    \hat \Pi^B\frac{1}{\hat N^B}\hat \Pi^B
    \rangle 
+ \sum_{i=1}^3 \langle ({\hat S}_{i}^A 
+{\hat S}_{i}^B )\rangle^2,
$ for (\ref{VAR1masha})
\item $RHS_{\Delta(\hat S^{AB})^2_{{new}}}
  = RHS_{\Delta(\hat S^{AB})^2} + \bigg(\sqrt{\langle \hat \Pi^{A^2}\rangle -\langle \hat{\vec{S}}^A\rangle^2} - 
\sqrt{\langle \hat \Pi^{B^2}\rangle - \langle \hat{\vec{S}}^B\rangle^2}\bigg)^2$ for \ref{NEWVAR_NEWSTOKES}  
\end{itemize}.

The formulas for probability of non-vacuum events go as follows:
\begin{equation}
\label{pi}
 \langle\hat\Pi^{A}\rangle_{add}  = 1 - (\sech^2{\Gamma})^{m_H + m_V + 2}
\end{equation}
and
\begin{equation}
 \langle\hat\Pi^{B}\rangle_{add} = 1 - \delta_{m_H,0}\delta_{m_V,0}(\sech^2{\Gamma})^{2},
\end{equation}
 
 For any $m_H$ and $m_V$ we have $\langle \hat S_1^X\rangle_{add} = \langle \hat S_2^X\rangle_{add} = 0$ for $X = A,B$. Moreover  if $m_H=m_V$ also $\langle \hat S_3^A\rangle_{add} = \langle \hat S_3^B\rangle_{add} = 0$. For all the rest of the cases we get
\begin{equation}
 \label{S}
 \langle\hat S_3^A\rangle_{add} = \frac{m_H - m_V}{m_H + m_V + 2}\Big(1 + 
  \frac{2\sech^2 \Gamma (m_H + m_V + \sech^2 \Gamma)}{(m_H + m_V)(m_H + m_V + 1)} \Big),
\end{equation}
\begin{equation}
 \langle\hat S_3^B\rangle_{add} = -\frac{(m_H - m_V)(1 - (\sech^2\Gamma)^{2(m_H + m_V +2)} )}{m_H + m_V + 2},
\end{equation}

\begin{table}[h]
\begin{tabular}{M{1cm}  M{1cm} M{7.4cm} M{3.8cm} M{3.8cm}}
\rowcolor{DarkGray}{}
&&&&\\[-3ex]
\rowcolor{DarkGray}{$m_H$} & $m_V$ & $ \langle\hat\Pi^{B}\frac{1}{\hat N_B} \hat\Pi^{B}\rangle_{add} $ & $ \langle \hat S_3^A \hat S_3^B \rangle_{add}$ & $ \langle \hat S_1^A \hat S_1^B \rangle_{add}$  \\ [1.5ex]
&&&&\\[-3ex]
1 & 0 & $\frac{1}{2}  \big(\sinh ^2\Gamma \left(\cosh ^2\Gamma+3\right) - 2 \log \left(\sech^2\Gamma\right)\big)\sech^6\Gamma$ & $\frac{1}{3} \big(3 \sech^6\Gamma-\sech^4\Gamma-\sech^2\Gamma-1\big)$   &  $\frac{1}{3} \big(3 \sech^6\Gamma-\sech^4\Gamma-\sech^2\Gamma-1\big)$ \\ [1.5ex]
\rowcolor{Gray}{} &&&&\\[-3ex]
\rowcolor{Gray}{1} & 1 & $\frac{1}{96} \sech^8\Gamma \big(87 \cosh (2 \Gamma )+12 \cosh (4 \Gamma )+\cosh (6 \Gamma )-96 \log \left(\sech^2\Gamma \right)-100\big)$ &  $\frac{1}{15} \big(6 \sech^8\Gamma-\sech^4\Gamma-2 \sech^2\Gamma-3\big)$ &  $\frac{2}{15} \big(6 \sech^8\Gamma-\sech^4\Gamma-2 \sech^2\Gamma-3\big)$ \\ [1.5ex]
\end{tabular}
\caption{Formulas needed to calculate $RHS$ for entanglement conditions  with normalized Stokes operators.
}
\label{table:Normalized_Stokes}
\end{table}

We were not  able to find a concise formulas for arbitrary $m_H$ and $m_V$  for other expressions we give  their specific values  of $m_H$  and $m_V \in \{0,1\}$ in the table
\ref{table:Normalized_Stokes}.

Fig \ref{2modeaddNStokes} refers to  conditions
(\ref{VAR1masha}) and (\ref{NEWVAR_NEWSTOKES}) for normalized Stokes operators and non-gaussian BSV. The more photons we add, the bigger is the gap
between $\kappa(\Gamma)$ for these conditions.

\section*{Acknowledgements}
This work is supported by  Foundation for Polish Science (FNP), IRAP project ICTQT, contract no. 2018/MAB/5, co-financed by EU  Smart Growth Operational Programme.
This work is supported by  Foundation for Polish Science (FNP), IRAP project ICTQT, contract no. 2018/MAB/5, co-financed by EU  Smart Growth Operational Programme.

\newpage
\bibliography{Variance}

\end{document}